\documentclass[letterpaper, 10 pt, conference]{ieeeconf}  

\IEEEoverridecommandlockouts                              

\usepackage{graphics} 
\usepackage{epsfig} 
\usepackage{mathptmx} 
\usepackage{times} 
\usepackage{amsmath} 
\usepackage{amssymb}  

 \usepackage{multirow}
 \usepackage[linesnumbered,ruled,vlined,Algorithm ]{algorithm }
\usepackage{algpseudocode}
\usepackage{hyperref}
\usepackage{array}
\usepackage{caption}
 \usepackage{subcaption}
\usepackage{enumerate}
\newtheorem{theorem}{Theorem}[section]

\newtheorem{lemma}[theorem]{Lemma}

 \usepackage{cite}
\title{\LARGE \bf Restricted Airspace Protection using  Multi-UAV Spatio-Temporal Multi-Task Allocation }

\author{Shridhar Velhal$^{1}$ and Suresh Sundaram$^{2}$
\thanks{$^{1}$Shridhar Velhal is a PhD student at Department of Aerospace Engineering, Indian Institute of Science, Bengaluru, India.
       {\tt\small velhalb@iisc.ac.in}}%
\thanks{$^{2}$Suresh Sundaram is an Associate Professor at Department of Aerospace Engineering, Indian Institute of Science, Bengaluru, India.
        {\tt\small vssuresh@iisc.ac.in}}%
}

\begin{document}
\maketitle
\thispagestyle{empty}
\pagestyle{empty}

\begin{abstract}

This paper addresses the problem of restricted airspace protection from invaders using the cooperative multi-UAV system. The objective is to detect and capture the invaders cooperatively by a team of homogeneous UAVs (called evaders) before invaders enter the restricted airspace. The problem of restricted airspace protection problem is formulated as a Multi-UAV Spatio-Temporal Multi-Task Allocation problem and is referred as MUST-MTA. The MUST-MTA problem is solved using a modified consensus-based bundled auction method. Here, the spatial and time constraints are handled by combining both spatial and temporal loss component. The solution identifies the sequence of spatial locations to be reached by the evader at specific time instants to neutralize the invaders. The performance of MUST-MTA with consensus approach is evaluated in a simulated environment. The Monte-Carlo simulation results clearly indicate the efficacy of the proposed approach in restricted airspace protection against intruders.
\end{abstract}
\begin{keywords}
  Restricted Airspace protection,  Multi-Task Allocation, Consensus Based Auction, Spatio-Temporal Task.
\end{keywords}

\section{INTRODUCTION}

Rapidly evolving technologies in autonomous unmanned aerial vehicles (UAV's) and associated developments in low-cost sensor have created a significant interest among researchers, in using them for various civil and military applications. Particularly, the autonomous aerial vehicles are often used for logistics \cite{kuru2019analysis}, medical \cite{rosser2018surgical}, agriculture \cite{mogili2018review}, security and surveillance \cite{harikumar2019mission}, \cite{harikumar2018multi}. The increase in the use of UAV in lower altitude introduces many challenges in privacy, safety and security \cite{solodov2018analyzing}. These UAV's may be flying over critical infrastructure such as nuclear facility, airport, chemical industries, ports and so on. Protecting restricted airspace from the UAV's physical attack is really a challenging problem. Detecting and responding to the UAV's invaders over a restricted airspace plays an important role. 

\begin{figure}[ t!]
     \centering
    \includegraphics[scale=0.2]{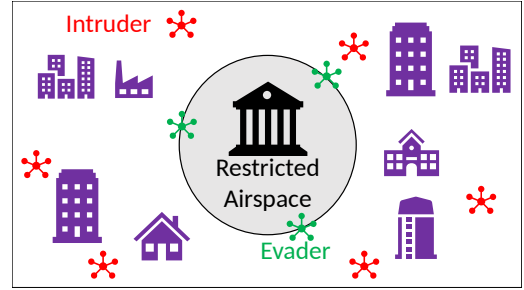}
    \caption{Restricted airspace protection using Multi-UAV system}
    \label{fig:TP1}
\end{figure}

First time in the literature, this paper address a cooperative multi-UAV system for restricted airspace protection from UAV invaders. In a typical Restricted Airspace Protection (RAP) problem as shown in Fig. \ref{fig:TP1}, the multiple evaders revolve around the airspace and cooperatively neutralize the multiple invaders moving towards the airspace. At any given time, evaders will be able to detect the invaders and estimate the time and location  of arrival in the region of engagement (spatio-temporal tasks). Note that these task are available only at a specific time and if these task are not handled by the evaders than invader will be able to attack the airspace successfully. Under a full communication scenario, the problem of RAP is formulated as a multiple-UAV spatio-temporal multi-task allocation problem. It is referred as MUST-MTA. The dynamically varying number of tasks, spatial and temporal dimensions adds to the complexity in cooperative task allocation between multiple evaders which minimize the overall effort. 

A composite loss function which computes the effort made by individual evaders to move in a specific sequence to reach the location of the task at a specific time instant. Note that the tasks assigned to the evader are unique. The problem of MUST-MTA is formulated as a linear integer programming and solved using modified consensus-based bundled algorithm (CBBA) proposed in \cite{Choi2009}. The algorithm utilizes market-driven decision strategy for decentralized multi-task allocation with time constraints and consensus routine to resolve conflict between the evaders. The algorithm allocates the spatio-temporal task which forms the path for the evaders in the region of engagement. The performance of the proposed MUST-MTA for RAP problem has been evaluated in a simulated environment. Further, Monte-Carlo simulation studies are carried-out to understand the effect of time-separation between the intruders on the point-of-failure.

The main contribution of the paper is: Formalization of the RAP problem into a multi-UAV spatio-temporal multi-task allocation problem. The presence of spatial and temporal dimensions and dynamic environments makes the solution for MUST-MTA challenging. A composite loss function is defined to handle the spatio-temporal nature of the task. The linear integer programming problem is solved using modified consensus based bundle allocation.

This paper is organized as follows. The related work is discussed in  \ref{sec:related_work}.  Section \ref{sec:MUST MTA} defines the RAP problem, spatio-temporal tasks,  formulate the MUST-MTA problem, and present the modified CBBA method. Section \ref{sec:Simulation_results} provides the simulation results of the restricted airspace protection problem. Finally, a conclusion is given in Section \ref{sec:Conclusion}
 
\section{Related Work} \label{sec:related_work}

One of the important challenges in the use of multi-UAV system for real-world applications is a complex task allocation problem between agents under unknown/uncertain environment. The objective in the complex task allocation is to find optimal strategy that will assign a set of tasks to the UAV such that multi-UAV system achieves its goal. More detailed review of task allocation and taxonomy of task allocation can be found in \cite{korsah2013comprehensive}, \cite{khamis2015multi}. Recent task allocation literature focus on dynamic allocation of spatially located tasks using market driven strategies \cite{jones2007learning}, game theoretic strategies \cite{Cui2013}, Hungarian method \cite{chopra2014heterogeneous,Chopra2017} and consensus based task allocation \cite{Choi2009,Brunet2008,zlot2006market,fanti2018decentralized}. 
Recently, in \cite{amador2014dynamic,nelke2020market} fisher market clearing based task allocation approach is presented to handle dynamically allocated spatial task which requires certain time to complete the task as a time-window constraints. The spatial task with time window constraint is solved using heuristic methods, where a penalty is imposed on delayed execution of tasks. More detail on existing algorithms on multi-task allocation with time-window constraints can be found in \cite{nunes2017taxonomy}.
The above-mentioned works are not suitable for RAP because the tasks are dynamic and are available only at a specific time instant. 

Issac et al \cite{wishart1966differential} introduced the concept of territory guarding differential game played by invaders and evaders. The goal of evader is to capture the invader as far as from the territory, and goal of invader is to avoid capturing and reach as close as possible towards the territory. Several research works are available in the literature to address territory guarding problem \cite{hsia1993first,lee2002strategy,analikwu2016reinforcement,raslan2016learning, analikwu2017multi}. These works either uses single invader and single evader or two evader to handle single invader. Recently, deep reinforcement learning based defensive escort team is proposed to avoid active collision \cite{garg2019defensive}.   Here, escort team tries to safely navigate the payload by positioning escort around the payload. (Payload co-relates to the restricted airspace in RAP.) As obstacles are repelled by escorting agents, agents will need to position them in a way such that they repel the obstacles on route cooperatively. Although these approaches handle dynamic tasks, the intruders are not directed as in RAP problem. Further, in RAP, evaders have to reach the spatial location at specific instant to neutralize the invaders. 

\section{ Restricted Airspace Protection Problem }    \label{sec:MUST MTA}

\subsection{Problem Definition} \label{sec:problemTP}
\begin{figure}[ t!]
    \centering                                \includegraphics[scale=0.45]{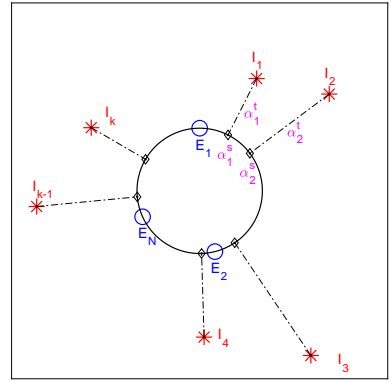}  
    \caption{Restricted Airspace (Territory) protection problem }
    \label{fig:TP2}
\end{figure}

The restricted airspace depends on the critical infrastructure; Any shaped infrastructure is approximated by circle. The scenario of RAP for a time instant t is shown in Fig. \ref{fig:TP2}. A cooperative multi-UAV team called as `evaders', will protect the airspace from intruders. The evaders will operate only inside the territory and neutralize the intruder in a ring around the restricted airspace referred to as `region of engagement'. One evader can neutralize an intruder when they come close within a neutralizing distance (r). 
Evaders will detect the positions of intruders. Depending upon the position information. Each evader needs to cooperatively decide for a sequence of tasks, following which it can neutralize invaders. 

The terms used in the RAP problem are as follows,
\begin{itemize}
    \item $Territory$: It is restricted airspace to be protected.
    \item $Evader$: A UAV which protects the restricted airspace.
    \item $Invader$: A UAV which tries to enter the restricted airspace.
    \item $Region \ of \ Engagement$: A region (ring) around the restricted airspace where invader can be neutralized. 
    \item $Task$: $\bf{\alpha} (\alpha^{s},\alpha^{t})$ : A task is defined by every intruder; intruder will penetrate from location $\alpha^{s}$ at time $\alpha^{t}$. 
    \item $Neutralizing \  point$ ($\alpha^{s}$): A location in RoE from where an intruder tries to enter.
    \item $Time \  of \  intrusion$ ($\alpha^{t}$):  A time  at which intruder  will enter restricted airspace if not neutralized.
    \item $Path$ ${\bf p}_i$: It is a sequence in which evader $i$ will execute the tasks. 
 \end{itemize}
 
 Evaders  ($E_1 , E_2,..E_N $) are less in number than  intruders ($I_1, I_2,..,I_K  $). Evaders will fly with maximum speed $v_E^{max}$.   Only one evader is sufficient to neutralise one intruder; more over once an intruder is neutralised, evader is free to do another task. The path ${\bf p}_i = \{ \alpha_x , \alpha_x, \alpha_z\}$ means the evader $i$ will  execute task $\alpha_x$,   $\alpha_y$, and  $\alpha_z$ sequentially.

The following assumptions are made for RAP problem,
\begin{enumerate}[{A}1)]   
    \item  \label{A_1} Each intruder $I_j$ is moving with a constant speed $v_j$, directed towards the centre of the restricted airspace. 
    \item \label{A_2} All evaders are homogeneous. Using same sensors they will identify the intruders position and velocity.  All evaders can communicate with each other.
    \item \label{A_3} The evader has higher velocity than that of intruder. 
    \item  \label{A_4} Intruders are not attacking at the same time.
    \item \label{A_5} All intruders and evaders are operating at the same height.
\end{enumerate}

 The assumption \hyperref[A_4]{A4} is very critical because evader will be helpless when a large number of the intruders approach at different locations at the same time instant. 

 Intruder $I_j$ will try to enter the airspace with a velocity of $v_j$, directed towards the centre.
 Evader has to neutralize the intruder $I_j$ at location $\alpha_j^{s}$ from where intruder tries to enter. 
 A task ${\bf{\alpha}}_j(\alpha_j^s , \alpha_j^t) $  is a 
 spatio-temporal task defined such that, evader should reach a location   $\alpha_j^s$  at a specific time  $\alpha_j^t$.
 

 
 \subsection{Spatio-temporal task} \label{sec:STT}
\begin{figure}[ hbt!]
    \centering
    \includegraphics[scale=0.5]{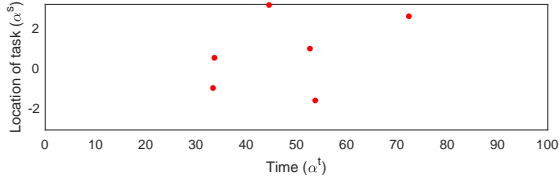}  
    \caption{Typical spatio-temporal tasks identified at a time instant }
    \label{fig:Spatio-temp}
\end{figure}

 A task ${\bf{\alpha}}_j$ is generated by each intruder $I_j$. The tasks ${\bf{\alpha}}_j$ consists of two dimensions namely, spatial and temporal. The spatial dimension is to reach a location $\alpha_j^s$ and temporal dimension is to complete the task at time $\alpha_j^t$.  
 Each task has to be executed at a specific location at a specific time. Hence task is at a distinct point on the graph with two dimensions: space and time.
 For clarity, spatial location is converted to angular position on region of engagement. 
 Figure \ref{fig:Spatio-temp} shows a typical spatio-temporal tasks identified at a time t.
 In general, spatio-temporal tasks have n-dimension of spatial location and one temporal dimension.
   

\begin{lemma} \label{lemma_1}
If the number of the intruder tries to enter territory at the same time with no radial separation, is more than the number of the evaders then  solution is infeasible.
\end{lemma}

Consider $N_i$ number of intruders are trying to intrude at time $t$. In spatio-temporal task, it is required to execute $N_i$  tasks at a time $t$. As the intruder are not radially separated, they will reach RoE at different location at same time instant. The number of locations to be reached at time $t$ are more than the number of evaders available; Hence, all intruders can not be neutralized. The solution for these cases is infeasible.


\subsection{Multi-UAV Spatio-Temporal Multi-Task Allocation } \label{sec:MUST}
 Each intruder $j$ generates a spatio-temporal task $\alpha_j$ and evaders have to execute all tasks cooperatively. As intruders are more than evaders, evader has to do multiple tasks.  The tasks are assigned to evaders by solving a MUST-MTA problem. Here, intruders are modelled as tasks and evaders are the agents who execute the task. In the rest of the paper, agents are denoted by $i,k$ and task is denoted by $j$.

Multi-task allocation is a decision making whether a task is assigned to an agent or not. Also, if tasks are sequential, then MTA solves for the sequence in which tasks to be done.  The Loss function is used to quantify the task, depending upon that tasks are assigned to agents. The spatio-temporal task  has both spatial and temporal loss function.
Composite loss function has been designed by uniting spatial and temporal loss functions.

\subsubsection{Spatial component \texorpdfstring{$L_{ij}^s({\bf p}_i)$ }{TEXT}}
It is the component defined for agent $i$, to execute spatial task $\alpha_j^s$ on path ${\bf p}_i$ .
\begin{equation} \label{eq:L_s}
     L_{ij}^s ( {\bf p}_i  ) =   {\|(  \alpha_j^s  -  E_i^p({\bf p}_i )  ) \|}_2   + \eta {\|(  \alpha_j^s  -  I_j   ) \|}_2
\end{equation}
 where, $\alpha_j^s$ is spacial requirement of task $j$,  $E_i^p({\bf p}_i)$ is the  location of previous task on path ${\bf p}_i$. For first task on path  $E_i^p = E_i $, the location of agent $i$, $\eta \in (0 ,1)$ is scaling factor.
 The first term  in Eq.\eqref{eq:L_s} is the effective distance travelled for reaching the spatial location along path $ {\bf p}_i $.  The second term in  Eq.\eqref{eq:L_s} is distance of intruder from neutralising point; this value is constant for each task independent of agent. 
 
 The effective distance travelled by an agent on path $\{A,B,C\}$ is computed sequentially. Effective distance for reaching $B$ is distance between $AB$. The effective distance to reach $C$ along path $\{A,B,C\}$ is the distance between $BC$ as the distance $AB$ is already considered for reaching $B$.

\subsubsection{ Temporal component \texorpdfstring{$L_{ij}^t$}{TEXT} }  \hfill
   Firstly we compute, time at which intruder $j$ enters RoE, reffered as time of intrusion ($alpha_j^t$),
 \begin{align}
     \alpha_j^t  &=  \frac{   {\|( I_j  - \alpha_j^s      ) \|}_2 }{v_j}
 \end{align}
 The temporal component  $ L_{ij}^t({\bf p}_i)$ is defined for agent $i$, to execute task $j$ at time $\alpha_j^t$ along path ${\bf p}_i$, $I_j$ is locatopn of intruder $j$
\begin{equation} \label{eq:L_t}
      L_{ij}^t ( {\bf p}_i) = \begin{cases}  ( 1 +  \alpha_j^t) \left(\alpha_j^t - \alpha_{j^p}^t({\bf p}_i )\right) &\text{if $j$ is feasible on ${\bf p}_i$ } \\
      \infty &  \text{if $j$ is infeasible }
      \end{cases}
 \end{equation}
 where $\alpha_{j^p}^t({\bf p}_i )$ is the time of arrival of previous task on  path ${\bf p}_i$; $\alpha_{j^p}^t = 0$ for first task on path. 
 
 The task  $j$ is feasible on path ${\bf p}_i$ if task is executed with positive time step, and spatially feasible. 
 The positive time step is mathematically represented as
 $\left( \alpha_j^t - \alpha_{j^p}^t({\bf p}_i )\right) > 0 $.   
  The spacial feasibility means  distance between evader and the spatial task, is reachable with in the temporal requirement of task; mathematically written as,  $$\frac{ {\|(  \alpha_j^s  -  E_i^p({\bf p}_i )  ) \|}_2 }{v_{max}^E} <  \left( \alpha_j^t - \alpha_{j^p}^t({\bf p}_i )\right) $$. 
\subsubsection{Composite loss function \texorpdfstring{$L_{ij} $ }{TEXT}} 
 The composite loss function is defined for a spatio-temporal task as,
 \begin{equation} \label{eq:composite_f}
       L_{ij}({\bf p}_i ) =  f \left( L_{ij}^s({\bf p}_i ) ,L_{ij}^t({\bf p}_i ) \right)
 \end{equation}
 The function $f$ can be any nonlinear function; here  $ f(x,y) = xy $ hence above Eq.\eqref{eq:composite_f}  reduce to
  \begin{equation}  \label{eq:L}
       L_{ij}({\bf p}_i ) =  L_{ij}^s({\bf p}_i )    L_{ij}^t({\bf p}_i )  
 \end{equation}
 Now, MTA is defined as an linear integer programming.

 \subsubsection{Multi-task allocation problem}
 The goal of the task allocation algorithm is to assign each of the $ N_t $ tasks to the available $N $ agents such that single task is assigned to only one agent. The cost associated with assigning a task $j$ to the agent $i$ is $c_{ij}$.  $\delta_{ij}$ is decision variable  for assigning agent $i$ to task $j$
 The task assignment problem is defined as 
 \begin{subequations} 
  \addtocounter{equation}{-1}
\begin{align} \label{eq:Integer_prog}
     \min_{\delta_{ij}}   \quad & \sum_{i=1}^{N_a} \left(\sum_{j=1}^{N_t} c_{ij} \delta_{ij}\right)  \\ 
 {\rm such \ that} \qquad  &\sum_{i=1}^{N_u} \delta_{ij} \le 1\qquad \forall j \in {\cal J}  \label{eq:cost_cond_1}   \\   
 & \delta_{ij} \in \{0,1\}\qquad \forall (i,j) \in {\cal I} \times {\cal J} \label{eq:cost_cond_2} 
\end{align}
\end{subequations}

the condition  Eq.\eqref{eq:cost_cond_1} enforces that the task can be assigned to only one agent. Eq. \eqref{eq:cost_cond_2} is decision variable either agent   $i$ is assigned to task $j$or not.

The objective is to find the sequence (path) assigned to individual evader such  that overall goal is achieved, The is cost of over all goal given below:
  \begin{align} \label{eq:cost}
    c_{ij}[{\bf p}_i] &= 
  \begin{cases} 
     \min_ { n \le  \vert {\bf p}_i\vert }  L_i^{{\bf p}_i \oplus_n \{j\} }  - L_i^{{\bf p}_i }& \text{if $ j \notin  {\bf p}_i $ } \\
     \infty   & \text{if $ j \in  {\bf p}_i $ }  
  \end{cases}    \\[2pt]
   {\rm where, } & \qquad
      L_i^{{\bf p}_i }  = \sum_j L_{ij}^{{\bf p}_i } 
 \end{align}
 $ \vert . \vert$ is the cardinality of path, and $\oplus_n  \{ j \} $ adds the $j$ after n$^{th}$ element. As the task $j$ is added at any location, the cost of the new task is the difference between new path cost and original path cost.
 
{\bf{Remark:}}   The loss function computed in equations \eqref{eq:L_s}, \eqref{eq:L_t}, and \eqref{eq:L} considers the path, but their computation is based on only previous task listed in path, and not on the complete path ${\bf p}_i$. Hence, for the cost of the path is computed by summing loss function of all tasks in ${\bf p}_i$. The cost of the newly added task is the difference in path cost due to the new task. 


\subsection{Modified Consensus-based bundled auction algorithm  } \label{sec:CBBA}

\begin{algorithm} [t!]
 \caption{modified CBBA for agent $i$ at iteration $q$  } \label{algo:auction}
 \begin{algorithmic}[1]
\State {\bf{procedure}}   input $ {\bf b}_i(q-1) $, ${\bf p}_i(q-1) $, $ {\bf y}_i(q-1) $, $ {\bf z}_i(q-1)$
 \State $ {\bf b}_i(q)  =   {\bf b}_i(q-1) $; \qquad $ {\bf p}_i(q)  =   {\bf p}_i(q-1) $ 
 \State $ {\bf y}_i(q)  =   {\bf y}_i(q-1) $; \qquad $ {\bf z}_i(q)  =   {\bf z}_i(q-1) $
 \State \text{conflict resolved}  $= 0$ \;
 \While {\text{conflict resolved} $= 0 $}
    \% {\textit{Auction Algorithm}   }  \; 
   \State    $c_{ij} = \min_ { n \le  \vert {\bf p}_i\vert }  L_i^{{\bf p}_i \oplus_n \{j\} }  - L_i^{{\bf p}_i }, \ \  \forall j \in {\cal J } \backslash {\bf b}_i $ \; 
   \State    $h_{ij} =   {\mathbb I}(c_{ij} < y_{ij}), \qquad \forall j \in {\cal J } $ \;
  \State $ J_i = argmin_j    \  c_{ij} . h_{ij} $ \;
  \State $  n_{i,J_i} = argmin_j    \  L_i^{{\bf p}_i  \oplus_n \{j\} } $ \;
   \State ${\bf b}_i  = {\bf b}_i  \oplus_{end} {J_i} $ \;
   \State ${\bf p}_i  = {\bf p}_i  \oplus_{n_{i,J_i}} {J_i} $ \;
   \State ${  y}_{i,J_i}(q) = c_{i,J_i} $\;
  \State ${  z}_{i,J_i} = i $ \;
 \State  \textbf{Call  Consensus Algorithm  }\;
 \EndWhile \; 
\end{algorithmic}
  (Remark: minimization over all $\infty$  value is taken as $\infty$.  All $\infty$ means that the task $j$ is infeasible along path ${\bf p}_i $) 
\end{algorithm}

  \begin{algorithm}[hbt!]
 \caption{Consensus by agent $i$ at iteration   $q$  } \label{algo:consensus}
 \begin{algorithmic}[1]
\State {\bf{procedure}}   input $ {\bf b} ^k$, ${\bf p}^k$, $ {\bf y} ^k$, $ {\bf z} ^k$  (data received from agent $k$ via synchronized communication )  $i \ne k$ , $\ m \neq \{i,k \}$ \;
 \If{ $z_{kj}^k = k  \ \&  \ z_{ij}^i = k $ }   
 \State Update 
 \EndIf
 \If { $z_{kj}^k = k \  \& \  y_{kj}  <    y_{ij} $} 
\State Update  
\EndIf
  \If{ $z_{kj}^k = i  \ \&  \   z_{ij}^i = k $ }
 \State Reset
 \EndIf
 
 \If{ ${  z}_{kj}^k = m  \  \&  \ {  z}_{ij}^i \neq m $  }
 \If { ${  y}_{mj} ( = y_{kj}) <    {  y}_{ij}   $} 
\State Reset
 \EndIf
\EndIf 
\If {  $  {\bf z}_i  = {\bf z}_k     \qquad  \forall i,k \in {\cal I }  $ } 
\State  \text{conflict resolved} $ = 1 $
\EndIf
\State Update : $  y_{ij} = y_{kj} , z_{ij} = z_{kj} $ \;
\State Reset : $ y_{ij} = \infty  ,  z_{ij} =  \emptyset $ \;
\end{algorithmic}
\end{algorithm}

The CBBA \cite{Choi2009}  is a distributed task allocation algorithm  in which agents bids for a bundle of tasks which would be executed in a specific path. Path ${\bf p}_i$ is the sequence in which agent $i$  is assigned the task. 
The composite for spatial temporal task $c_{ij}({\bf p}_i)$ is the cost associated with agent $i$ doing task $j$ in path  ${\bf p}_i$.  The formulated cost function is minimized in CBBA. 
In CBBA, bids which win the task (smallest bidding value for task $j$) are denoted by ${{\bf y}_i }$, the wining agent's list is denoted by ${{\bf z}_i }$. $ {\bf p}_i $ is the path of performing tasks and $ {\bf b}_i $ bundle of tasks. The path $ {\bf p}_i $ gives the sequence in which tasks will be executed by agent $i$.

The Algorithm is initialized, as winning value of cost ${\bf y}_i = \infty $ and allocation  $ {\bf z}_i =  {\bf 0}  $. The bundle and path are initialized as null, $ {\bf p}_i ={\bf b}_i = \emptyset $.  
Agents bids independently in the auction irrespective of other agents, hence computation of bidding is distributed. The conflicts among the agents are resolved by consensus; which requires communication for exchanging information of bidding value, winning value, and path.

The algorithm \ref{algo:auction} presents the steps for auction. In which each agent bids for feasible targets independently. During this procedure, they also fix their path of execution of the task. 
Each agent bids greedily for all feasible tasks; this bidding information is shared with all agents  via synchronous communication. As bidding is performed independently, there may be conflicts between agents for the task. Consensus obtained for resolving these conflicts as given in algorithm \ref{algo:consensus}.
In consensus, a task is given to the lowest cost valued agent for that task, and this will be updated in the path, bundle, and winning agent vectors. If a new task is added to the path, the task is added at a location,  where it minimizes the path cost. When a task is removed from the agent then the entire bundle needs to be removed. The removal of a task changes the path of that agent, as other tasks are selected based on the path which is no longer valid; the entire path needs to be cancelled.

\section{Simulation Results and Discussion}    \label{sec:Simulation_results}
\begin{figure*}[t]
\centering
\begin{subfigure}{0.25\linewidth}
\includegraphics[width=1\linewidth]{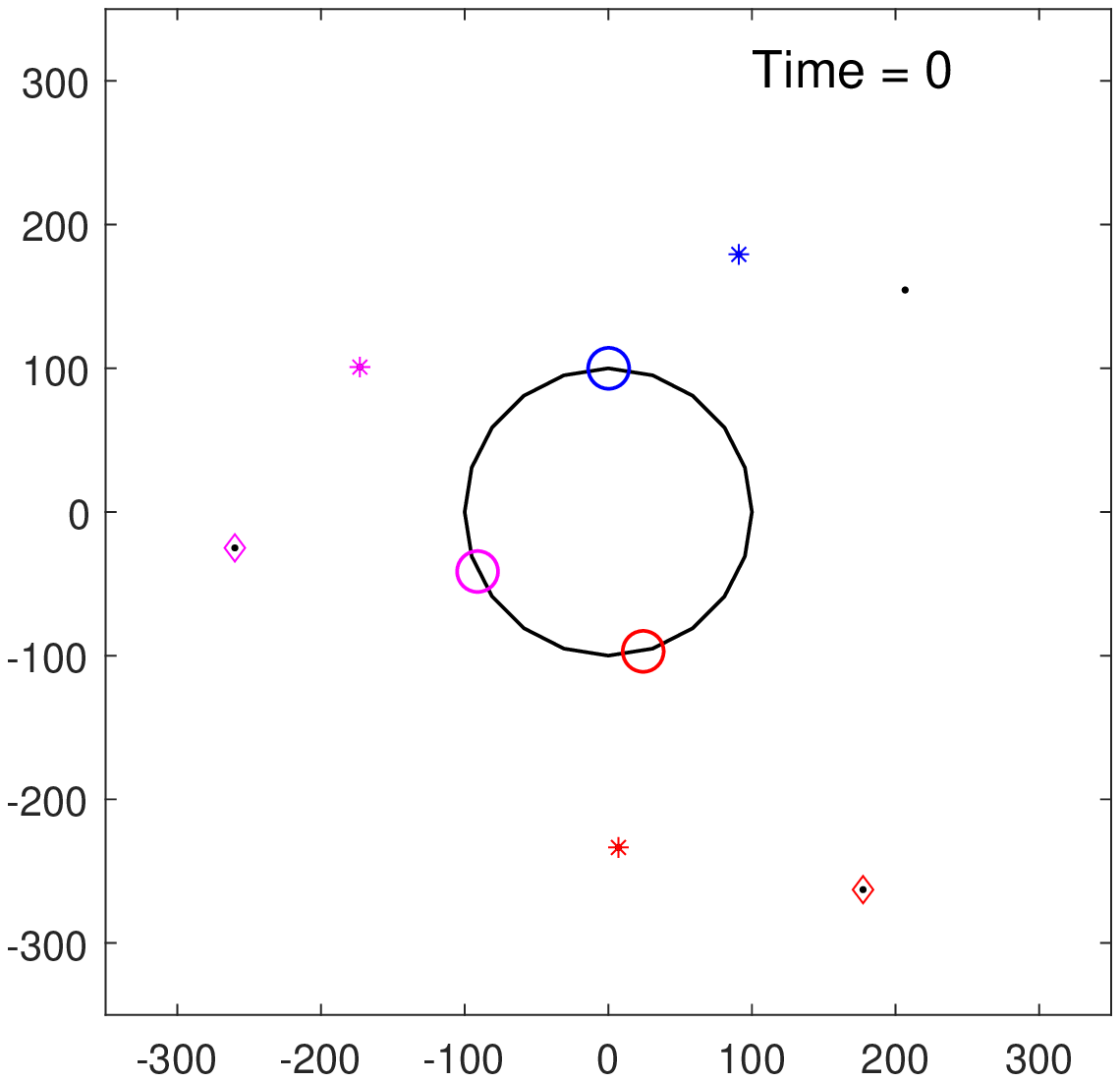} 
\caption{} \ \ 
\label{fig:subfig1}
\end{subfigure}%
\begin{subfigure}{0.25\linewidth}
\includegraphics[width=1\linewidth]{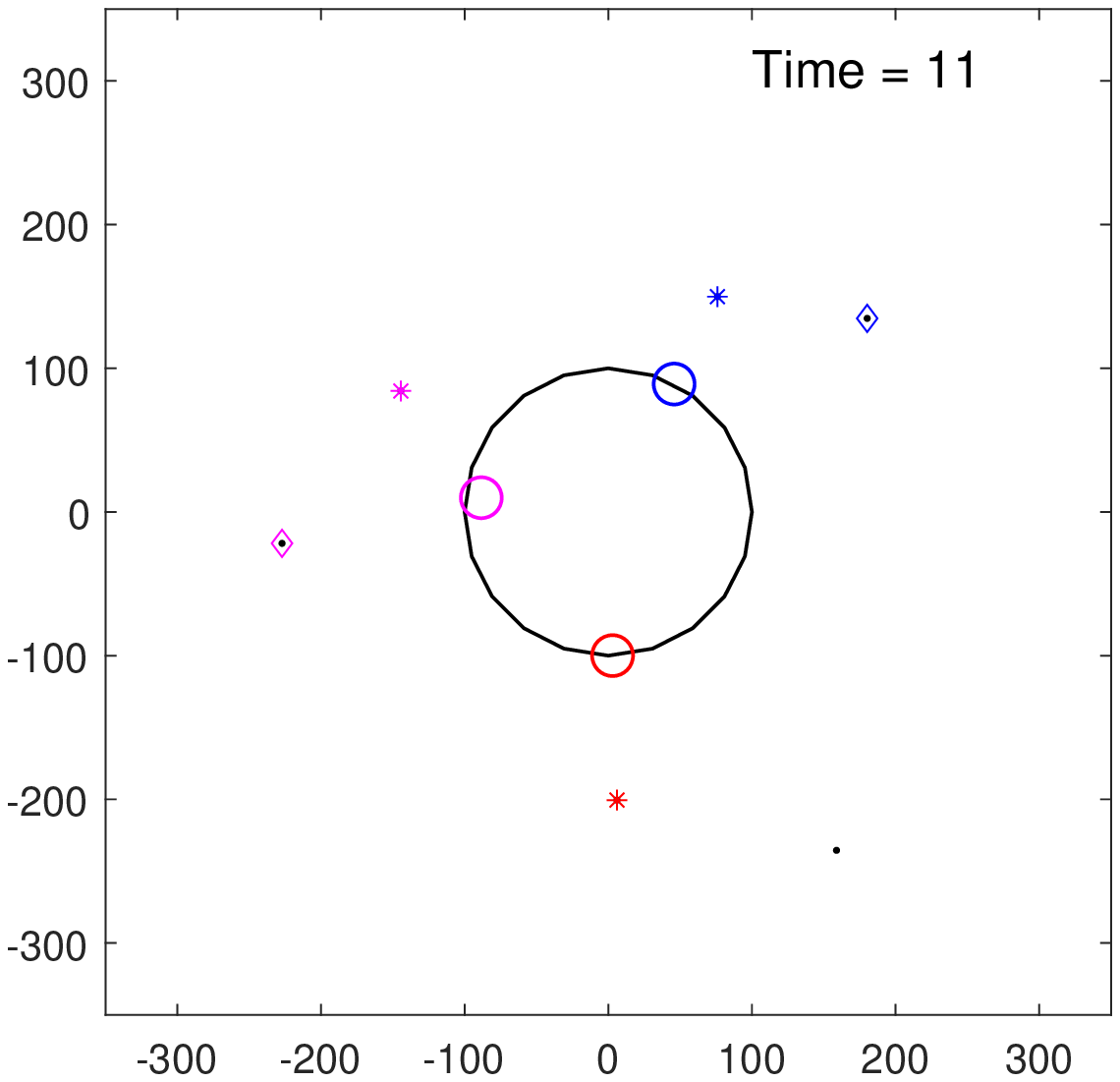}
\caption{} \ \ 
\label{fig:subfig2}
\end{subfigure}%
\begin{subfigure}{0.25\linewidth}
\includegraphics[width=1\linewidth]{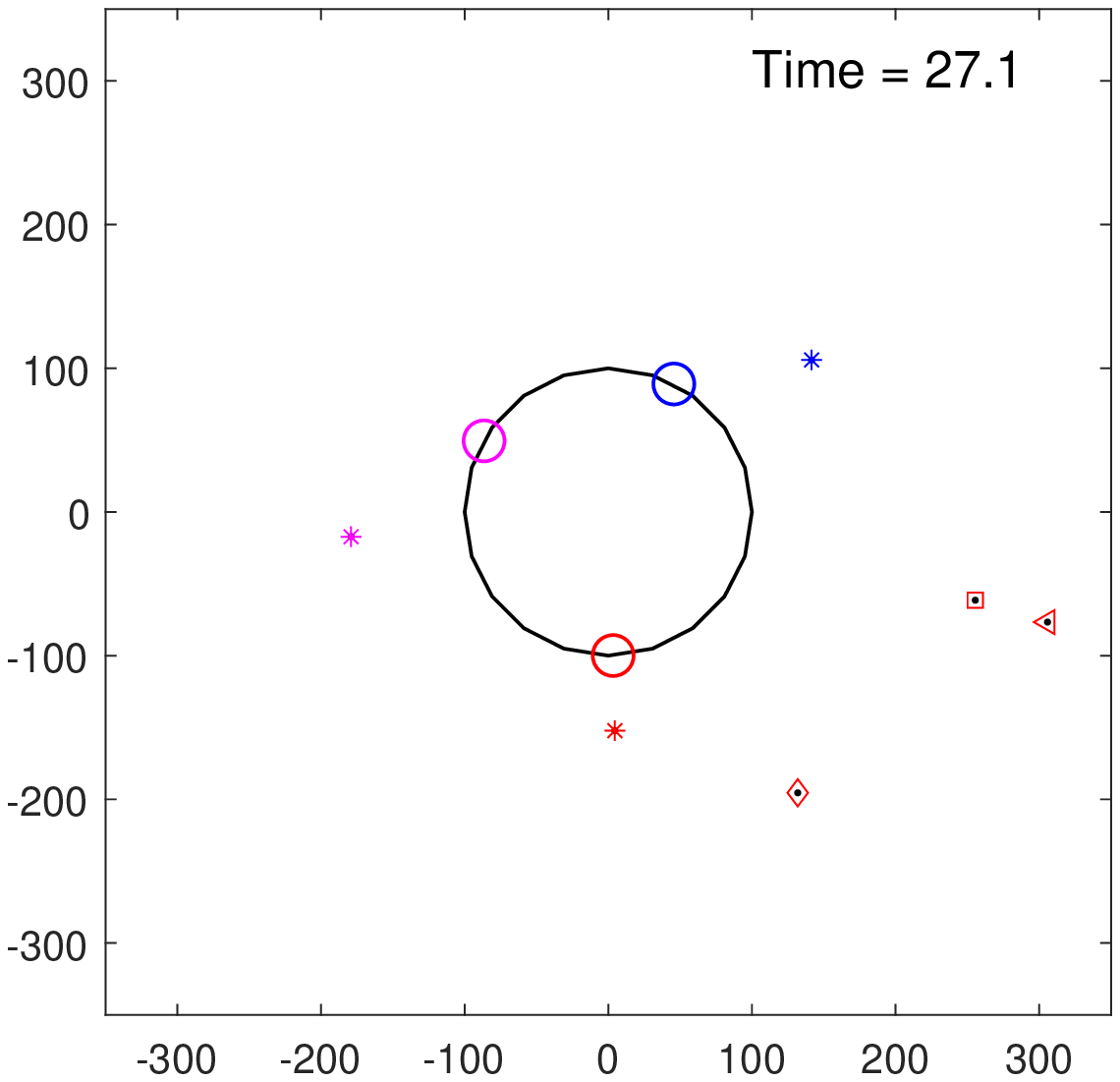} 
\caption{} \ \ 
\label{fig:subfig3}
\end{subfigure}%
\begin{subfigure}{0.25\linewidth}
\includegraphics[width=1\linewidth]{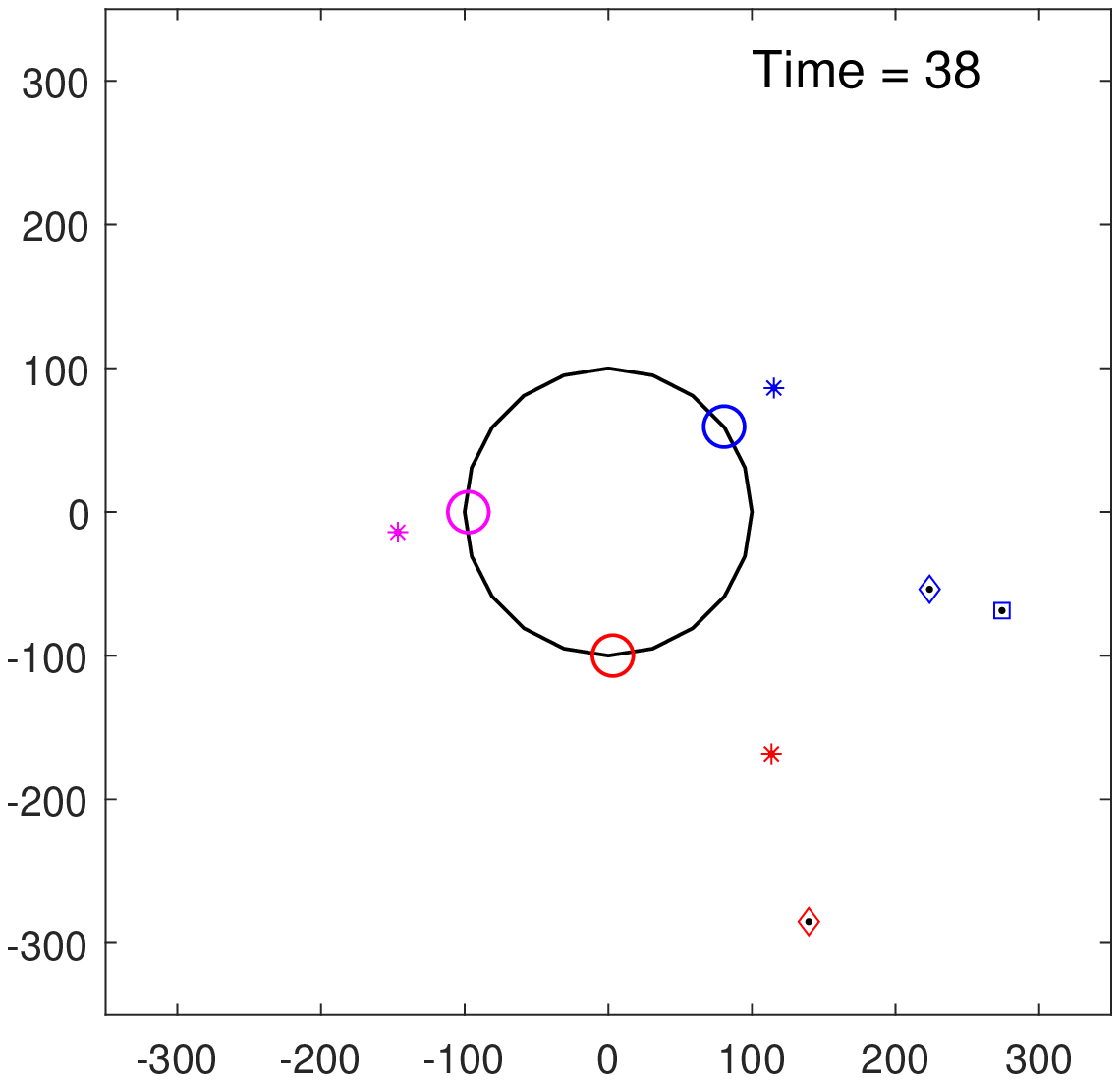}
\caption{} \ \ 
\label{fig:subfig4}
\end{subfigure}
\vspace{2pt}

\begin{subfigure}{0.25\linewidth}
\includegraphics[width=1\linewidth]{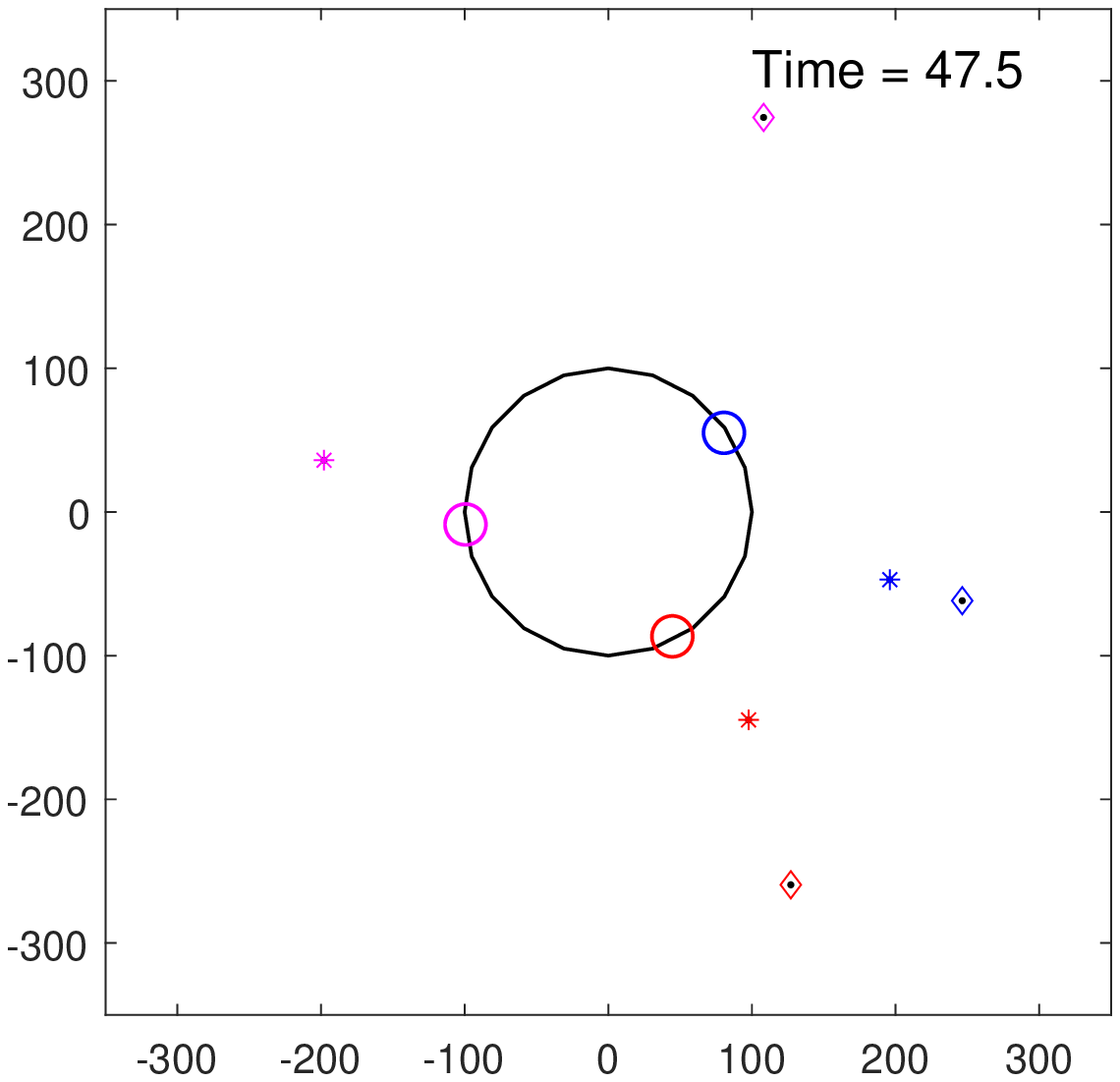} 
\caption{} \ \ 
\label{fig:subfig5}
\end{subfigure}%
\begin{subfigure}{0.25\linewidth}
\includegraphics[width=1\linewidth]{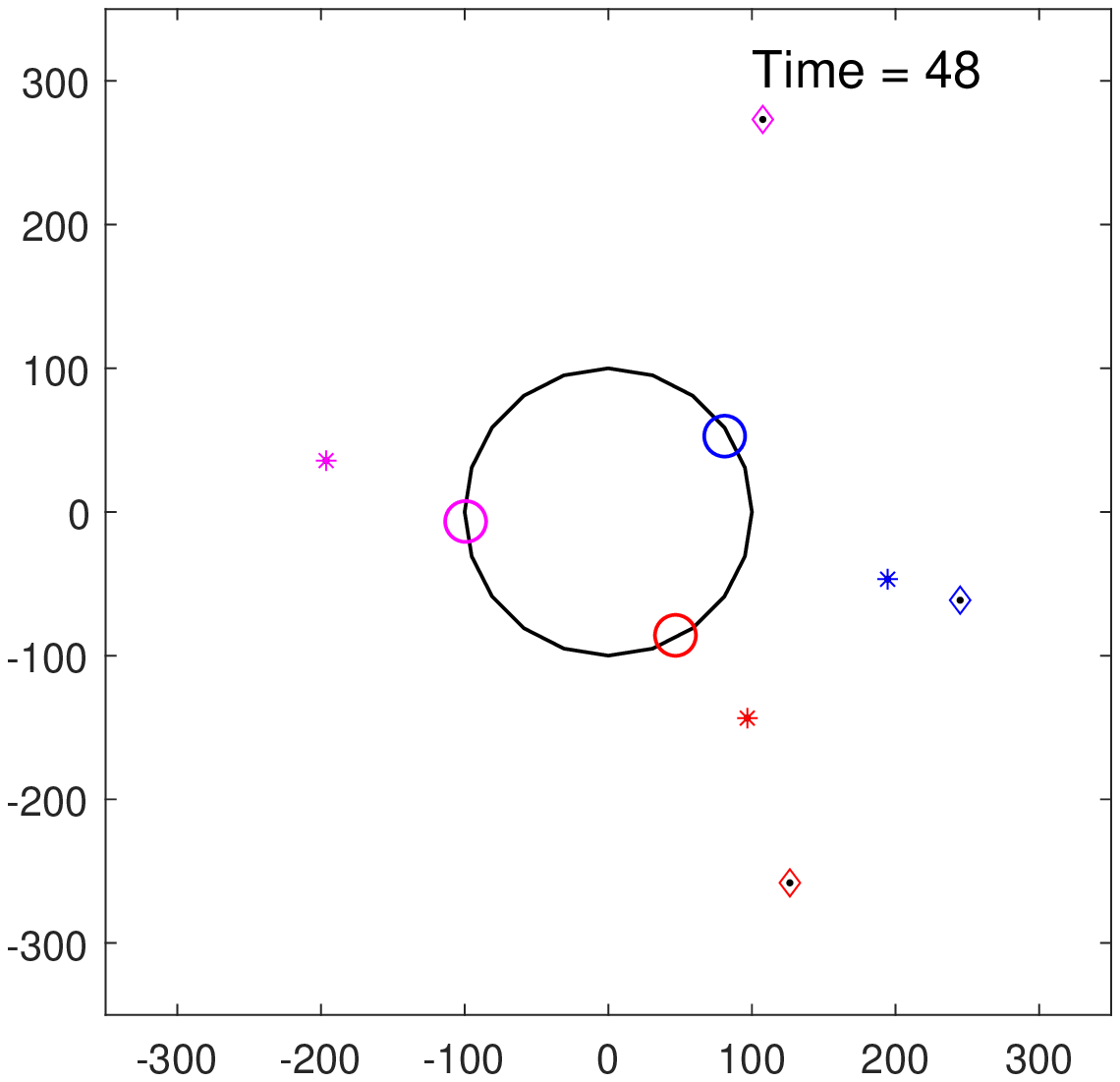}
\caption{} \ \ 
\label{fig:subfig6}
\end{subfigure}%
\begin{subfigure}{0.25\linewidth}
\includegraphics[width=1\linewidth]{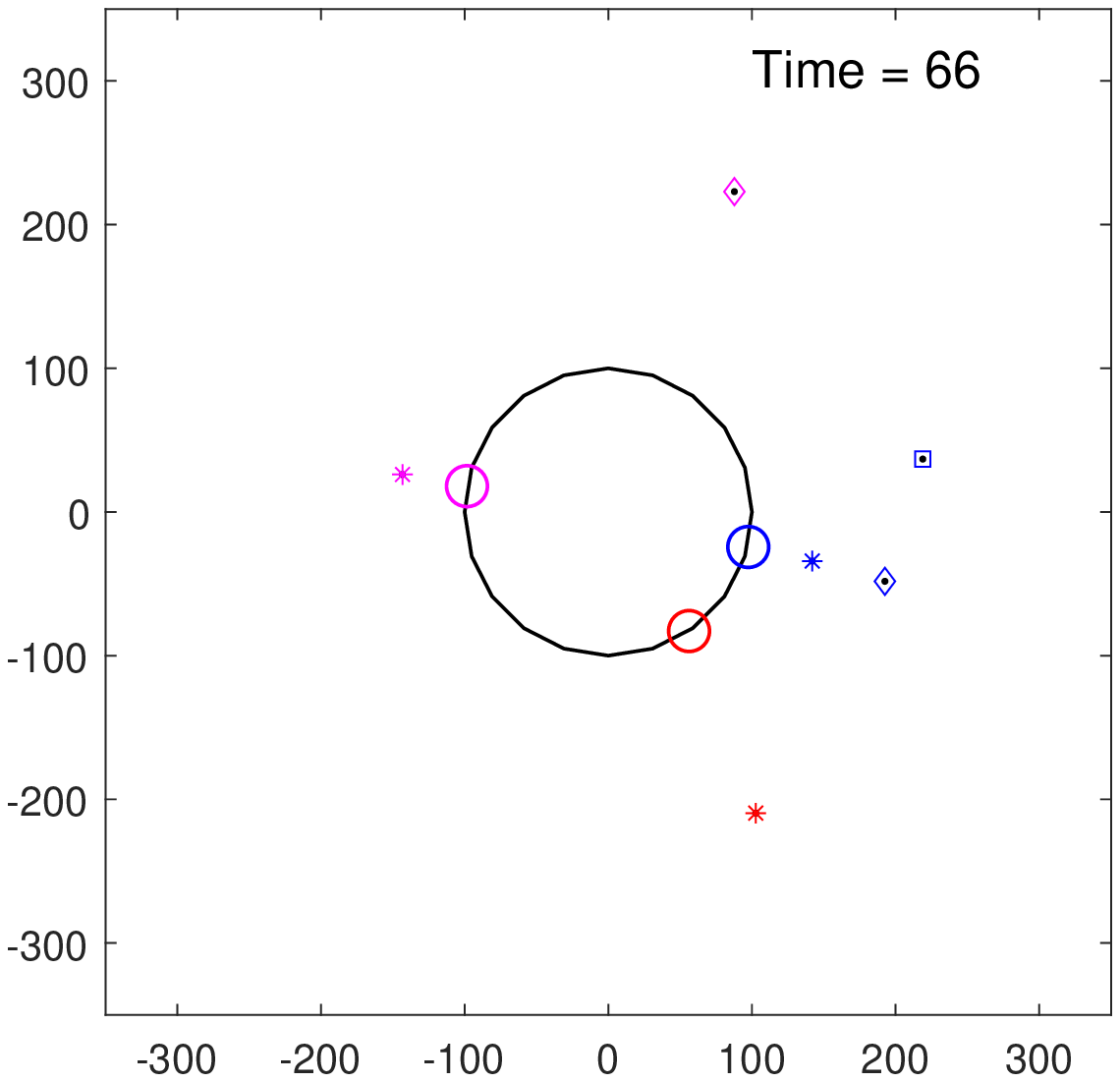} 
\caption{} \ \ 
\label{fig:subfig7}
\end{subfigure}%
\begin{subfigure}{0.25\linewidth}
\includegraphics[width=1\linewidth]{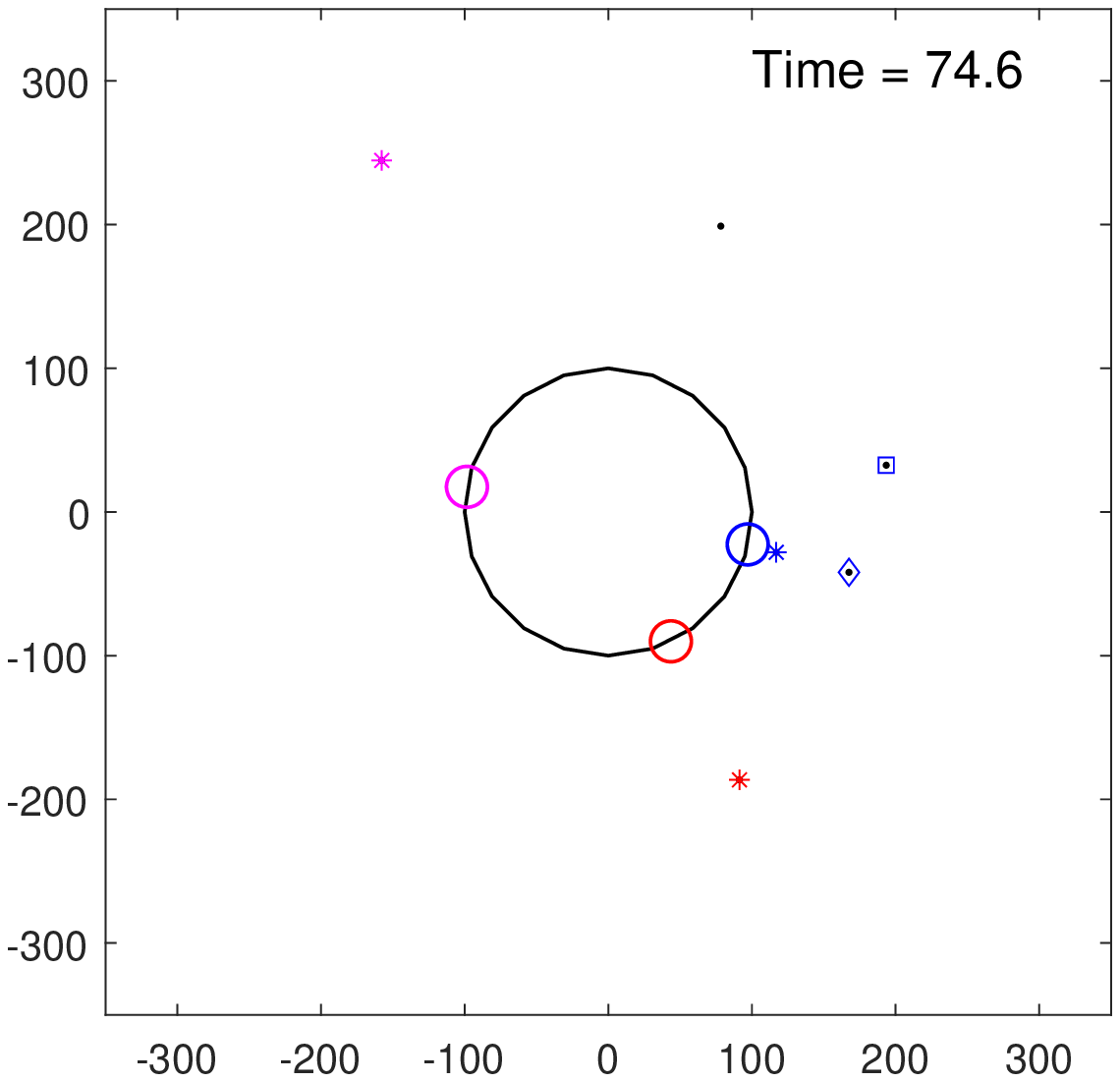}
\caption{} \ \ 
\label{fig:subfig8}
\end{subfigure}
\vspace{2pt}

\begin{subfigure}{0.25\linewidth}
\includegraphics[width=1\linewidth]{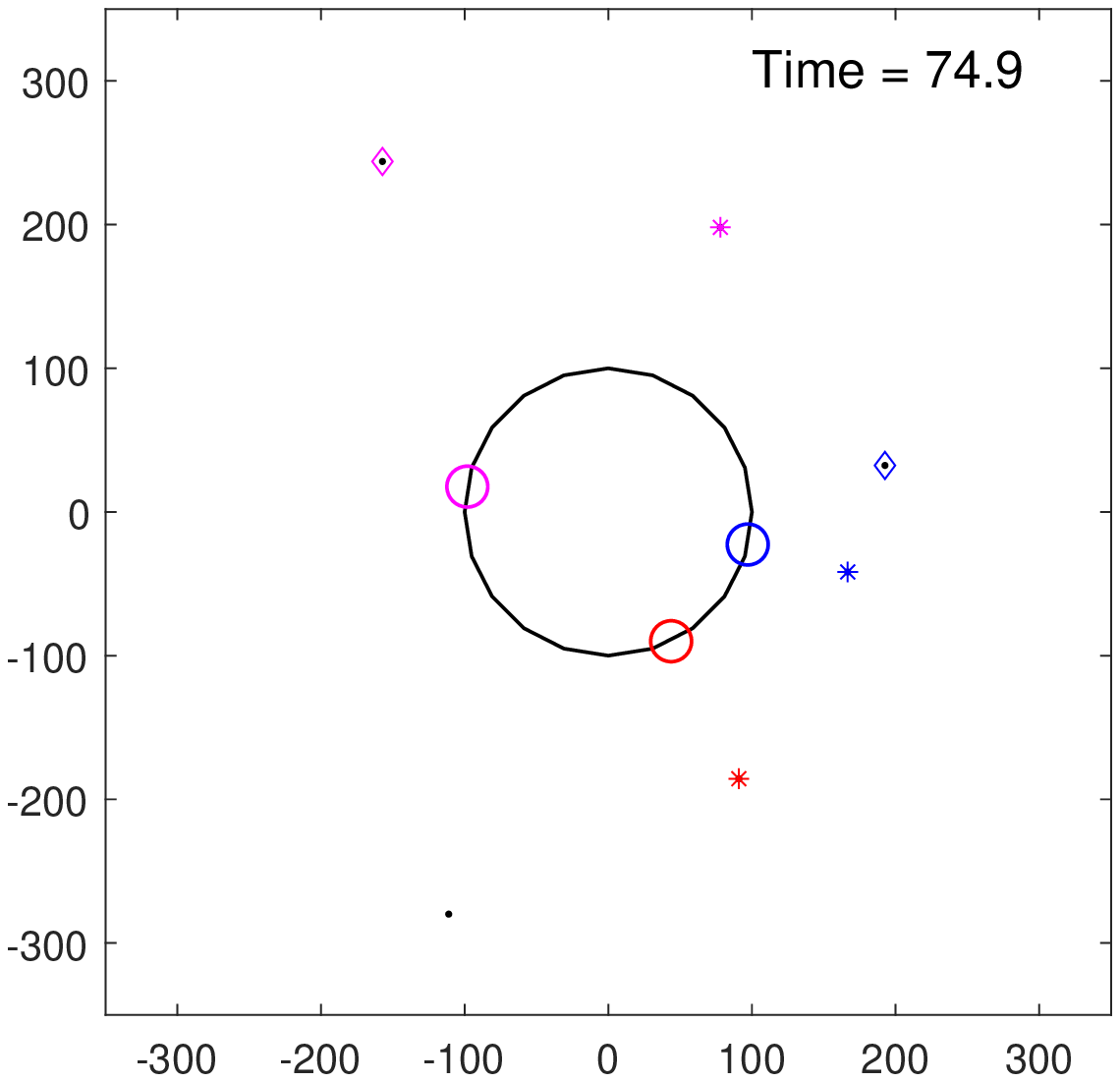} 
\caption{} \ \ 
\label{fig:subfig9}
\end{subfigure}%
\begin{subfigure}{0.25\linewidth}
\includegraphics[width=1\linewidth]{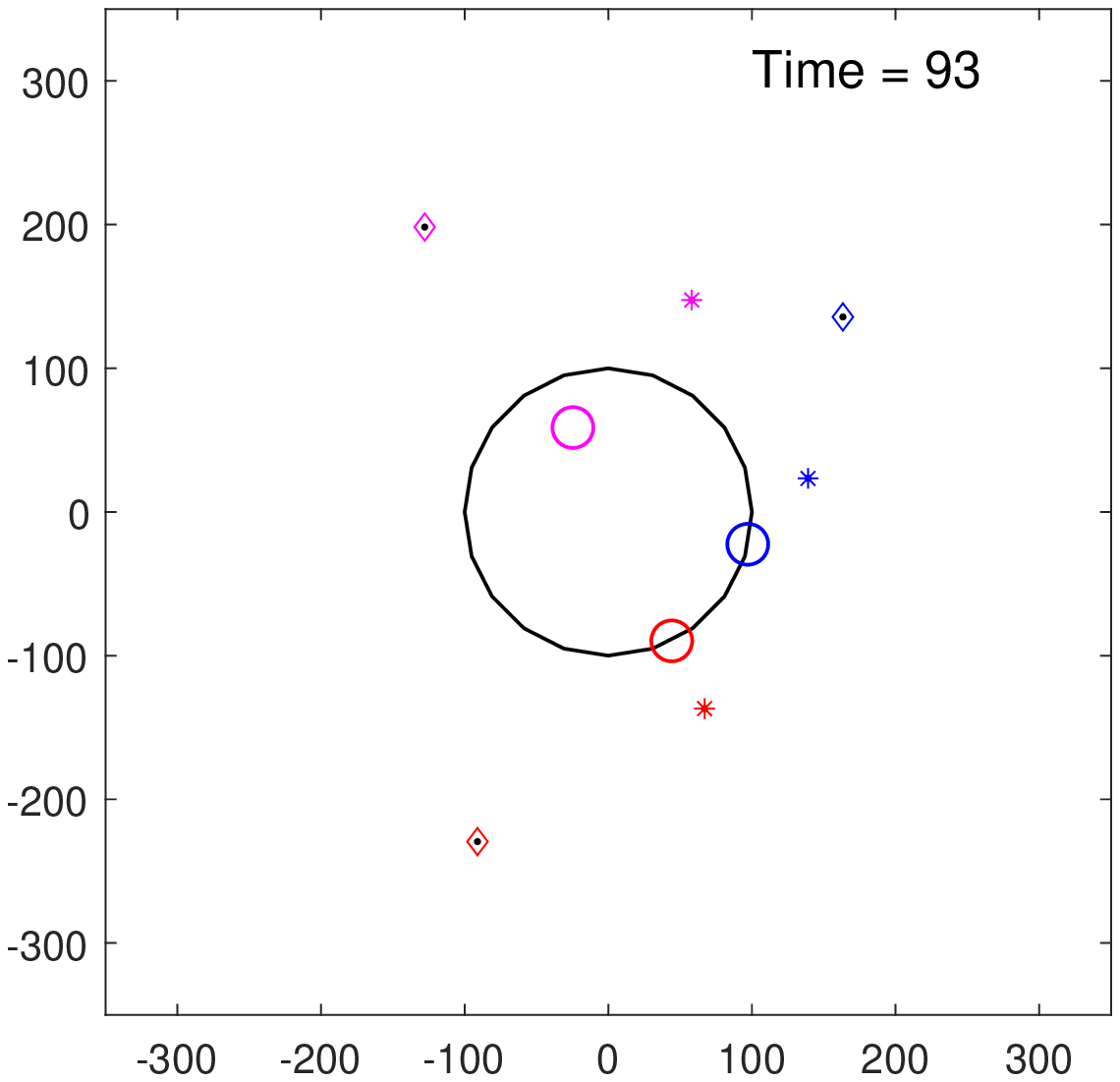}
\caption{} \ \ 
\label{fig:subfig10}
\end{subfigure}%
\begin{subfigure}{0.25\linewidth}
\includegraphics[width=1\linewidth]{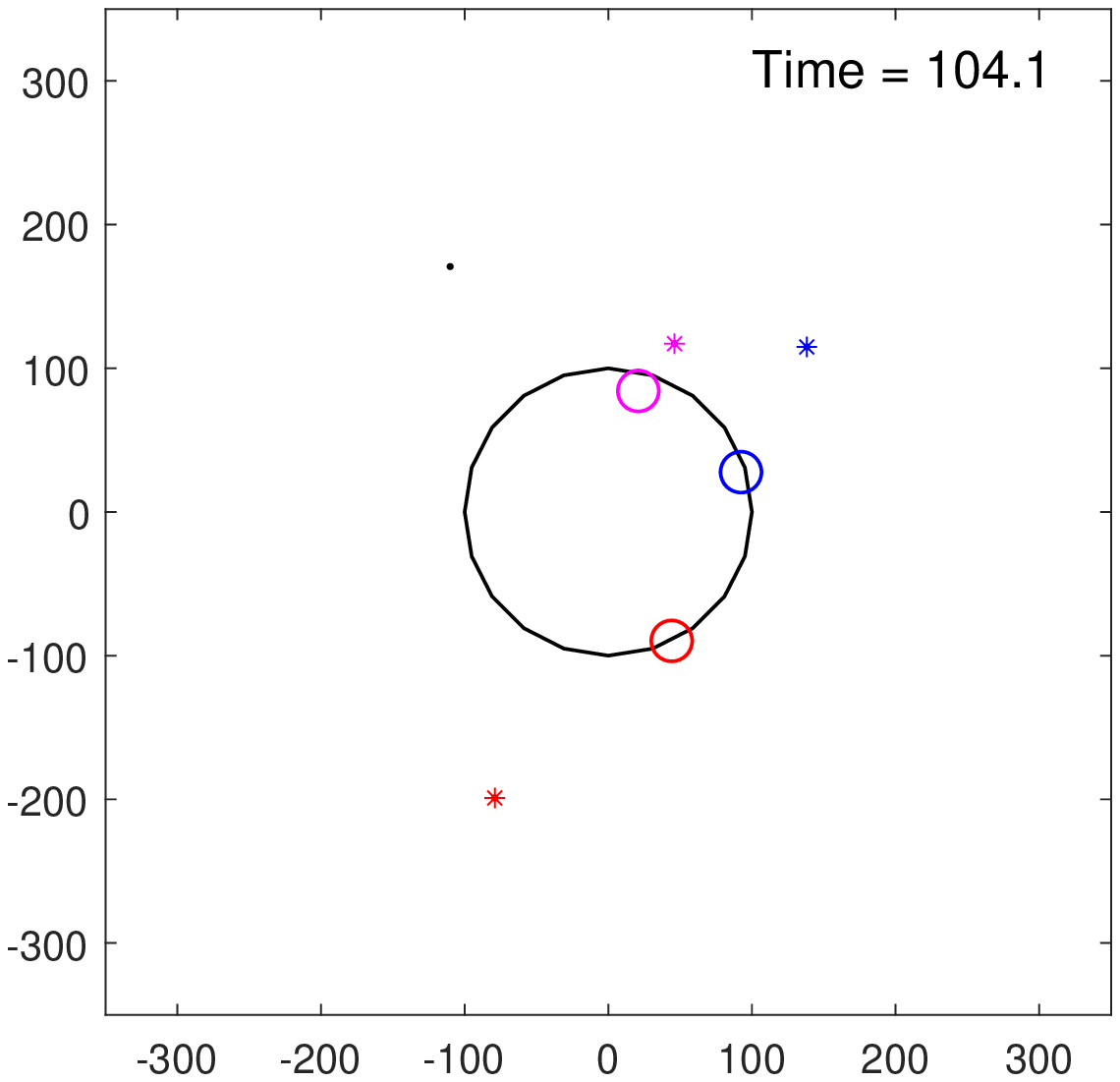} 
\caption{} \ \ 
\label{fig:subfig11}
\end{subfigure}%
\begin{subfigure}{0.25\linewidth}
\includegraphics[width=1\linewidth]{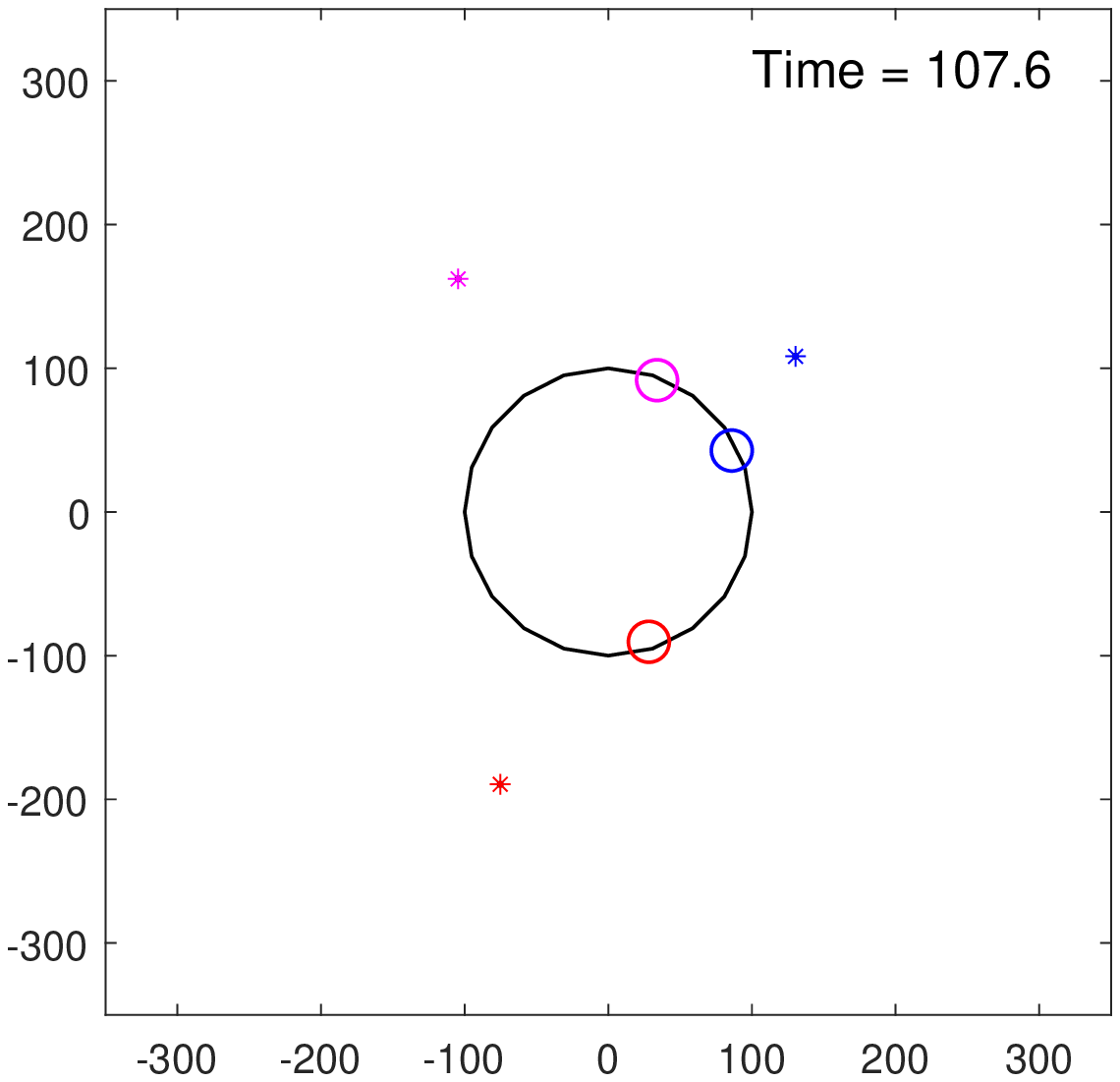}
\caption{} \ \ 
\label{fig:subfig12}
\end{subfigure}
\caption{Snapshots of the intruder tasks allocated to the evaders, at different time instants
\textmd{The evaders  $E_1,E_2,E_3$ are represented by colours  blue, red,  magenta respectively. } 
\textmd{The intruders assigned to an evader are coloured with the colour of evader.} 
\textmd{The path selected by evader is shown by symbols in sequence :  *,  $\diamond $, $ \square $, $\lhd$, $\rhd$, $\otimes$ }
 } 
 \label{fig:result_1}
 \end{figure*}

The simulations are performed for RAP problem. The restricted airspace is selected as a circle of radius of $ R = 100 $ m, The number of the evaders is  $K  = 3 $ and the number of the intruder is set  $N_i = 6$.  The evader will neutralise an intruder within neutralising distance  $ (r)$ set to  $20$ m. The velocity of all intruder is constant and selected as $v_I = 3$ m/s. The maximum velocity of all evaders  is  constrained by $ {v}_E^{max} = 4.5$ m/s.  When intruder is neutralised, new intruder is added at random location, in the neighbouring region of territory. The motion of evader and invader are computed using kinematic equations. This simulation are conducted in MATLAB R2019b in windows 10 environment.

\subsection{Case study and discussion}

\begin{figure}[ b!]
    \centering
    \includegraphics[scale=0.45]{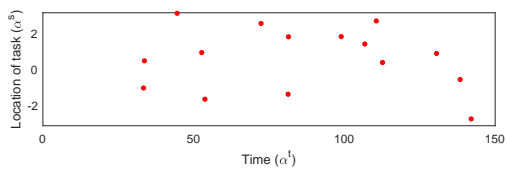} 
    \caption{Spatio-temporal tasks used for simulation case study}
    \label{fig:Spatio-temp_result}
\end{figure}

Fig. \ref{fig:Spatio-temp_result} shows the spatio-temporal tasks for test scenario discussed in Fig. \ref{fig:result_1}. Total 15 spatio-temporal are avilable during time interbval of $0-150$ sec The temporal tasks are well separated, and hence all tasks can be executed, and this has been shown in results. The temporal separation plays a critical role in the feasibility of the task. As tasks are temporally separated so all tasks are executed successfully and shown in \ref{fig:result_1}

Fig \ref{fig:result_1} shows the screenshots the task assignment computed  for a test scenario of RAP problem. Total 15 intruders are neutralised in this case, in which intruders are added at random locations. The task allocation algorithm is solved at every sampling time instant. The path of agents are shown in Fig. \ref{fig:result_1}, at different time instants. Figure \ref{fig:subfig1} shows the initial position of intruder and the   evaders. The intruders are allocated to all three agents, but one intruder is unassigned. One intruder is unassigned because that task is infeasible with the path chosen by agents. All agents move with the time that task may become feasible as observed in \ref{fig:subfig2}.

The agents $E_1$ and $E_3$ neutralise their respective first intruder; new intruders  from the random location is approaching towards the terrain as shown in Fig. \ref{fig:subfig3}. Furthermore, agent $E_1$ and $E_3$ are assigned to one task each while, agent $E_2$ is assigned to four tasks. Since it is cooperative task different number of task by agents.   In Fig. \ref{fig:subfig4}, it is observed that previous task allocation gets reallocated by the entry of a new intruder. The tasks of agent $E_2$ are allocated to agent $E_1$. Reallocation is because of two reasons; agent $E_2$  moves away from task due to newly added intruder; $E_2$ changes path, and the task becomes infeasible in a new path. Fig \ref{fig:subfig8} shows that task on the top right is unassigned, this is due to that fact that the task is feasible by agents $E_1$, but agent $E_1$  has another task which is less costly than that task. However, after a few time steps, task is taken by agent $E_3$ \ref{fig:subfig9}.  This happens due to sub-optimality in CBBA \cite{Choi2009}.

\begin{figure}[ t!]
    \centering
    \includegraphics[scale=0.4]{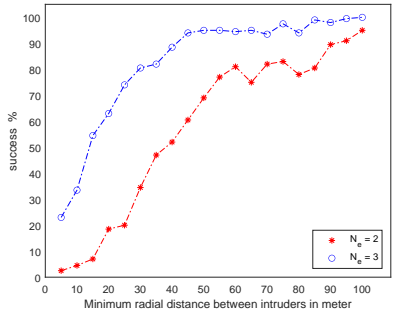}  
    \caption{Monte-Carlo simulations by varying radial distance between intruders and varying number of evader}
    \label{fig:MC_result}
\end{figure}

Monte-Carlo simulation is performed by varying the radial distance between intruders. As velocity of the intruder is constant the radial distance provides a temporal separation between tasks. A randomized simulations with 200 epoch are carried for a different radial distance for a newly added intruder. For an epoch, maximum of 30 intruders are added. When an intruder penetrates the restricted airspace, RAP fails. Figure. \ref{fig:MC_result} shows the percentage of success rate with different minimum radial distances; as the intruders gets more radial separation the success percentage increases. For the low value of radial separation, the tasks become infeasible as explained in Lemma \ref{lemma_1}. 
The number of the evaders are varied to study the minimum evaders required a given restricted airspace. It is expected that more number of the evaders can execute for less separation  tasks  The less number of evaders protecting restricted airspace requires more temporal separation between tasks.
 \section{Conclusion}    \label{sec:Conclusion}
Restricted airspace protection (RAP) from UAV intruders using the cooperative multi-UAV system is addressed. The movement of intruder towards the restricted airspace leads to the spatio-temporal task (capturing the intruder) for the multi-UAV (evaders) system. RAP problem has been formulated as a multi-UAV spatio-temporal multi-task allocation (MUST-MTA) problem. This paper provides a multi-task allocation solution to a spatio-temporal task using modified consensus-based bundle algorithm. The cost function is modified via composite loss function, which unites spatial and temporal components. The simulation validates the efficacy of the proposed method. Simulated results show that the tasks can be executed if they have sufficient temporal separation. As the number of the evaders increases, the temporal separation requirement decreases gradually.

\bibliographystyle{IEEEtran}
\bibliography{main_bib.bib}

   \end{document}